# Title Page

## Applications of artificial intelligence in the analysis of histopathology images of gliomas: a review

Jan-Philipp Redlich[1,*], Friedrich Feuerhake[2,3], Joachim Weis[4], Nadine S. Schaadt[2], Sarah Teuber-Hanselmann[5], Christoph Buck[6], Sabine Luttmann[7], Andrea Eberle[7], Stefan Nikolin[4], Arno Appenzeller[8], Andreas Portmann[9], André Homeyer[1]

[1]Fraunhofer Institute for Digital Medicine MEVIS, Max-von-Laue-Straße 2, 28359 Bremen, Germany
[2]Hannover Medical School, Carl-Neuberg-Straße 1, 30625 Hannover, Germany
[3]Institute of Neuropathology, Medical Center - University of Freiburg, Breisacherstraße 64, 79106 Freiburg, Germany
[4]Institute of Neuropathology, RWTH Aachen University Hospital, Pauwelsstrasse 30, 52074 Aachen, Germany
[5]Department of Neuropathology, Center for Pathology, Klinikum Bremen-Mitte, Sankt-Jürgen-Straße 1, 28205 Bremen, Germany
[6]Leibniz Institute for Prevention Research and Epidemiology - BIPS, Achterstraße 30, 28359 Bremen, Germany
[7]Bremen Cancer Registry, Leibniz Institute for Prevention Research and Epidemiology - BIPS, Achterstraße 30, 28359 Bremen, Germany
[8]Fraunhofer Institute of Optronics, System Technologies and Image Exploitation IOSB, Fraunhoferstraße 1, 76131 Karlsruhe, Germany
[9]German Heart Center Berlin, Augustenburger Platz 1, 13353 Berlin, Germany

*Corresponding author. E-mail address: jan-philipp.redlich@mevis.fraunhofer.de

# Abstract

In recent years, the diagnosis of gliomas has become increasingly complex. Analysis of glioma histopathology images using artificial intelligence (AI) offers new opportunities to support diagnosis and outcome prediction. To give an overview of the current state of research, this review examines 83 publicly available research studies that have proposed AI-based methods for whole-slide histopathology images of human gliomas, covering the diagnostic tasks of subtyping (23/83), grading (27/83), molecular marker prediction (20/83), and survival prediction (29/83). All studies were reviewed with regard to methodological aspects as well as clinical applicability. It was found that the focus of current research is the assessment of hematoxylin and eosin-stained tissue sections of adult-type diffuse gliomas. The majority of studies (52/83) are based on the publicly available glioblastoma and low-grade glioma datasets from The Cancer Genome Atlas (TCGA) and only a few studies employed other datasets in isolation (16/83) or in addition to the TCGA datasets (15/83). Current approaches mostly rely on convolutional neural networks (63/83) for analyzing tissue at 20x magnification (35/83). A new field of research is the integration of clinical data, omics data, or magnetic resonance imaging (29/83). So far, AI-based methods have achieved promising results, but are not yet used in real clinical settings. Future work should focus on the independent validation of methods on larger, multi-site datasets with high-quality and up-to-date clinical and molecular pathology annotations to demonstrate routine applicability.

# Introduction

Gliomas account for roughly a quarter of all primary non-malignant and malignant central nervous system (CNS) tumors, and 81% of all malignant primary CNS tumors[1]. Generally, gliomas can be differentiated into circumscribed and diffuse gliomas[2,3]. Whereas circumscribed gliomas are usually benign and potentially curable with complete surgical resection, diffuse gliomas are characterized by diffuse infiltration of tumor cells in the brain parenchyma and are mostly not curable, with an inherent tendency for malignant progression and recurrence[2,3]. In particular, glioblastoma, the most common and most aggressive malignant CNS tumor, makes up 60% of all diagnosed gliomas and caused 131,036 deaths in the US between 2001 and 2019 alone[1]. Although the introduction of now well-established standard therapy protocols, involving surgical resection, radiation, and chemotherapy[4] as well as guidance by MGMT promoter methylation status[5], has substantially improved clinical outcomes, 5-year survival in patients with glioblastoma has been largely constant at only 6.9%[1].

The current international standard for the diagnosis of gliomas is the fifth edition of the WHO Classification of Tumors of the Central Nervous System, published in 2021[6]. Based on advances in the understanding of CNS tumors, the WHO classification system has been constantly evolving since its first issue in 1979, with new versions released every 7–9 years[7]. Whereas previous versions greatly relied on histological assessment via light microscopy, the current classification system incorporates both histopathological features and molecular alterations (Figure 1), allowing for a more uniform delineation of disease entities[6,7]. The basic principle of classification according to histological typing, histogenesis, and grading, however, remains fundamental and underlines the role of conventional histological assessment as an important element of timely, standardized, and globally consistent diagnostic workflows[6].

The rapidly evolving field of computational pathology, in particular, the analysis of digital whole-slide images (WSIs) of tumor tissue sections using artificial intelligence (AI), offers new opportunities to support current diagnostic workflows. Promising applications range from automation of time-consuming routine tasks, such as subtyping and grading, to tasks that cannot be accurately performed by human observers, such as predicting molecular markers or survival directly from hematoxylin and eosin (H&E)-stained routine sections of formalin-fixed and paraffin-embedded (FFPE) tumor tissue[8–10]. Essential concepts of AI and AI-based analysis of WSIs are briefly explained in Tables 1 and 2 and Figure 2.

This review provides a comprehensive overview of the current state of research on AI-based analysis, including both traditional machine learning (ML) and deep learning (DL), of whole-slide histopathology images of human gliomas. Motivated by a significant increase in corresponding publications (Figure 3), this review focuses on the diagnostic tasks of subtyping, grading, molecular marker prediction, and survival prediction. Studies primarily addressing image segmentation[11–15], image retrieval[16], or tumor heterogeneity[17–22] have been considered out of scope. The reviewed studies are examined with regard to diagnostic tasks and methodological aspects of WSI processing, as well as discussed by addressing limitations and future directions. Prior related reviews on AI-based glioma assessment addressed other imaging modalities, such as computed tomography or magnetic resonance imaging (MRI)[23–29], and specific diagnostic tasks, such as grading[30,31] or survival prediction[32–34].

# Results

## Diagnostic tasks

The reviewed studies covered the diagnostic tasks of subtyping (23/83), grading (27/83), molecular marker prediction (20/83), and survival prediction (29/83). All studies focused on the assessment of adult-type diffuse gliomas (Figure 1) and nearly all were based on H&E-stained tissue sections (only two studies considered other histological staining techniques[35,36]).

The majority of the studies (52/83) were based on two public datasets from The Cancer Genome Atlas, which originated from the Glioblastoma[37,38] (TCGA-GBM) and Low-Grade Glioma[39] (TCGA-LGG) projects. Both contain information on diagnosis, molecular alterations, and survival for approximately 600 and 500 patients, respectively. Patients in both datasets had been diagnosed before 2013 and 2011[37–39] and therefore according to versions of the WHO classification system before 2016. The TCGA-GBM dataset includes primary glioblastomas (WHO grade IV), and the TCGA-LGG dataset includes astrocytomas, anaplastic astrocytomas, oligodendrogliomas, oligoastrocytomas, and anaplastic oligoastrocytomas (all WHO grade II or III). An independent research project later annotated regions of interest (ROIs) of relevant tumor tissue for both datasets and made these publicly available[40].

Other studies employed other datasets in isolation (16/83) or in addition to the TCGA datasets (15/83). Diagnoses in these datasets corresponded either to the current version of the WHO classification from 2021[41–43], or previous versions from 2016[44–46] or earlier. Table 3 presents an overview of all publicly available datasets utilized by the reviewed studies.

It should be noted that the 2021 WHO Classification of Tumors of the Central Nervous System changed the notation of tumor grades from Roman to Arabic numerals and endorsed the use of the term “CNS WHO grade”[6]. Throughout this review, these changes in notation are used only for diagnoses that were reported according to the 2021 WHO classification system.

### Subtyping

Subtyping gliomas is of fundamental importance for the diagnosis and subsequent treatment of patients[6,7]. Before 2016, the WHO classification system incorporated only histological features. Oligodendrogliomas, for instance, were characterized by features including round nuclei and a “fried egg” cellular appearance; astrocytomas by more irregular nuclei with clumped chromatin, eosinophilic cytoplasm, and branched cytoplasmic processes; and glioblastomas by an astrocytic phenotype combined with the presence of necrosis and/or microvascular proliferation[7]. The latest revisions of the WHO classification system published in 2016 and 2021 reflect important advances in the understanding of the molecular pathogenesis of gliomas by incorporating molecular alterations in an integrated diagnosis (Figure 1). For instance, since 2021, the diagnosis of glioblastoma has been limited to malignant non-IDH-mutated (i.e., IDH-wildtype) gliomas. Furthermore, the diagnosis of oligoastrocytomas has been removed from the 2021 WHO classification system. Until 2016, oligoastrocytomas were considered a distinct tumor entity with mixed oligodendroglial and astrocytic histological features, and between 2016 and 2021, they could only be diagnosed in cases with unknown 1p/19q and IDH status (these cases were designated as “oligoastrocytoma, NOS (Not Otherwise Specified)” and diagnosis was not recommended unless molecular testing was unavailable)[6,7]. Table 4 provides an overview of all corresponding studies grouped according to the investigated subtyping tasks.

Whereas one study aligned IHC- (for IDH1-R132H and ATRX) and H&E-stained tissue sections to differentiate gliomas along IDH mutation status and astrocytic lineage[35], all other studies predicted glioma types only from H&E-stained tissue. Prediction was usually performed in an end-to-end manner, i.e., without prior identification of established histological or molecular markers. Many approaches were based on weakly-supervised learning (WSL), in which predictions of patch-level convolutional neural networks (CNNs) were aggregated to patient-level diagnosis[46–49]. Aggregation of patch-level to slide-level results was either performed by majority voting[46,49] or traditional ML methods fitted to histograms of patch-level predictions[47,48].

Two studies focused on the analysis of intraoperative frozen sections (IFSs): Shi et al. (2023) proposed two weakly supervised CNNs for diagnosing intracranial germinomas, oligodendrogliomas, and low-grade astrocytomas from H&E-stained and intraoperative frozen sections (IFSs), respectively[46]. They compared the diagnostic accuracy of three pathologists with and without the assistance of their proposed models and reported an average improvement of 40% and 20%, respectively. Nasrallah et al. (2023) used a hierarchical Vision Transformer (ViT) architecture and three glioma patient cohorts to support several diagnostic tasks during surgery, including the identification of malignant cells, as well as the prediction of histological grades, glioma types according to the 2021 WHO classification system, and numerous molecular markers[42].

Three other recent studies also leveraged emerging ViT architectures for subtyping[41,45,50]. Notably, Li et al. (2023), employed TCGA data and a proprietary patient cohort for three tasks of brain tumor classification, including glioma subtyping, as well as prediction of IDH1, TP53, and MGMT status in gliomas[45]. Their proposed model featured an ImageNet pre-trained ViT for encoding image patches and a graph- and Transformer-based aggregator for modeling relations among patches. It achieved high performance in all tasks and automatically learned histopathological features, including blood vessels, hemorrhage, calcification, and necrosis.

Some studies aimed to improve the accuracy of subtyping by integration of MRI. A particularly large number of such studies emerged from the CPM RadPath Challenges 2018, 2019 and 2020, which provided paired magnetic resonance and whole-slide image datasets for the differentiation of astrocytomas, oligodendrogliomas and glioblastomas[13,51,52]. The best performing approaches used weakly-supervised CNNs (or an ensemble thereof[52]) for processing WSIs and fused WSI- and MRI-derived predictions by an average pooling[52], a max pooling[13] or by favoring WSI-derived predictions[51]. The improvement in balanced accuracy compared to using WSI alone was up to 7.8%[52]. Related to the integration of MRI, Mallya et al. (2022) utilized the same CPM-RadPath dataset and proposed a knowledge distillation framework to transfer the knowledge from WSIs to a model that differentiates gliomas based solely on MRI[53].

In addition to the aforementioned end-to-end approaches, some studies first predicted established histological or molecular markers, and then derived glioma types based on these predictions[41,50,54–56]. In a series of studies, Ma et al. (2023), for instance, proposed a classification model for the differentiation of astrocytoma, oligodendroglioma, ependymoma, lymphoma, metastasis, and healthy tissue, based on the presence of respective cell types, and glioma grades based on the presence of microvascular proliferation and necrosis[48,55]. They later assessed their model's performance using a multi-center dataset comprising 127 WSIs from four institutions[56]. Interestingly, Hewitt et al. (2023) compared approaches for predicting the 2021 WHO glioma subtypes, and reported that first predicting IDH, 1p/19q, and ATRX status, and then inferring subtypes from these predictions performed overall better than an end-to-end approach[41]. In addition, they conducted further interesting experiments, such as comparing the predictability of 2016 versus 2021 WHO glioma subtypes, and validated all results on independent patient cohorts.

## Grading

Complementing tumor types, tumor grades differentiate gliomas according to predicted clinical behavior and range from CNS WHO grade 1 to 4, with CNS WHO grade 1 indicating the least malignant behavior[2,6]. Similar to subtyping, grading is currently based on both histological and molecular markers (Figure 1), whereas in the past only histological markers, including cellular pleomorphism, proliferative activity, necrosis, and microvascular proliferation were considered[7]. Table 5 provides an overview of all corresponding studies grouped according to the investigated grading tasks.

All studies predicted glioma grades from WSIs of H&E-stained tissue sections in an end-to-end manner. Exceptions from this include one earlier study that inferred glioma grade from detected necrosis and microvascular proliferation[57], and a recent study that analyzed human leukocyte antigen (HLA)-stained tissue microarrays and inferred glioma grade by assessing the infiltration of myeloid cells in the tumor microenvironment[36].

Earlier studies mostly employed handcrafted features and traditional ML methods[15,57–62]. In a typical study, Barker et al. (2016) proposed a coarse-to-fine analysis based on handcrafted features extracted from segmented nuclei. In this manner, they first selected discriminative patches and subsequently predicted grades using a linear regression model[61].

All other studies employed CNNs for predicting grades from image patches. Three such approaches decomposed II vs. III vs. IV grading as stepwise binary II and III vs. IV followed by II vs. III classifications and agreed on greater difficulty in differentiating grades II vs. III[63–65]. Su et al. (2023) focused on this challenge and reported considerable performance improvement by employing an ensemble of 14 weakly-supervised CNN classifiers whose predictions were aggregated by a logistic regression model[66]. Whereas this approach seems computationally intensive, Momeni et al. (2018) proposed a deep recurrent attention model, for II and III vs. IV grading, which achieved state-of-the-art results while only analyzing 144 image patches per WSI[67].

Two recent studies examined the newly adopted approach of grading gliomas within types rather than across types, as endorsed by the 2021 WHO classification system[43,56]. Most notably, Wang et al. (2023) proposed a clustering-based CNN model for this purpose (similar to that of Zhao et al.[68]), which they validated on two external cohorts of 305 and 328 patients, respectively[43].

Several studies integrated additional modalities such as genomics[63,69,70], age and sex[60] or proliferation index manually assessed from Ki-67 staining[58]. Most notably, Qiu et al. (2023) adopted a self-training strategy to address the effects of label noise and proposed an attention-based feature guidance aiming to capture bidirectional interactions between WSIs and genomic features[69]. They demonstrated the superiority of multi-modal over uni-modal predictions (AUC 0.807 (WSI), 0.804 (genomics), 0.872 (WSI+genomics)) as well as the effectiveness of modeling bidirectional interactions between modalities for glioma grading and lung cancer subtyping.

Addressing the problem of possibly limited availability of genomics in clinical practice, Xing et al. (2022) proposed a knowledge distillation framework to transfer the privileged knowledge of a teacher model, trained on WSIs and genomics, to a student model which subsequently predicts grade solely from WSIs[71]. Remarkably, they reported superior performance of their uni-modal student model over a state-of-the-art multi-modal model[70] on the same validation cohort.

## Molecular marker prediction

Molecular alterations in gliomas have diagnostic, predictive, and prognostic value critical for effective treatment selection, e.g., the presence of MGMT promoter methylation was associated with a beneficial sensitivity to alkylating agent chemotherapy, resulting in prolonged survival in glioblastoma patients[2,5]. Standards of practice for accurate, reliable detection of molecular markers include immunohistochemistry, sequencing, or fluorescent in situ hybridization (FISH)[2]. Table 6 provides an overview of all corresponding studies grouped according to the investigated molecular markers.

Although there are no established criteria for determining molecular markers from histology yet, a growing body of research has suggested their predictability from WSIs of H&E-stained tissue sections with considerable accuracy. Among others, pan-cancer studies predicted various molecular markers across multiple cancer types[72–74] and with regard to glioblastoma reported an externally validated accuracy of over 0.7 for 4 out of 9 investigated markers[72].

Most studies predicted IDH mutation status. Similar approaches based on weakly-supervised CNNs were augmented by either the incorporation of additional synthetically generated image patches (AUC 0.927 vs. 0.920)[75] or by the integration of MRI (ACC from WSI: 0.860; from MRI: 0.780; from both: 0.900)[76]. As patient age and IDH mutation status are highly correlated (e.g., the median age at diagnosis was reported to be 37 years for IDH-mutant astrocytomas and 65 years for IDH-wildtype astrocytomas[1]), integration of patient age into multi-modal predictions further improved performance[75–77].

Comparing their performance in predicting IDH status to pathologists, Liechty et al. (2022) proposed a multi-magnification ensemble that averaged predictions from multiple weakly-supervised CNN classifiers, each processing individual magnification levels[78]. On an external validation cohort of 174 WSIs, their model did not surpass human performance (AUC 0.881 vs. 0.901), but averaged predictions from pathologists and their model performed on par with the consensus of two pathologists (AUC 0.921 vs. 0.920).

Recently, Zhao et al. (2024) proposed a clustering-based hybrid CNN-ViT model to predict IDH mutation in diffuse glioma[68]. They trained and assessed their model based on two large and independent patient cohorts from two hospitals, and, motivated by the changes in the 2021 WHO classification system, investigated subgroups with particularly challenging morphologies. Of note, their study provided a clear explanation of how the ground truth IDH status was determined through sequencing for all patients in both cohorts.

Five studies were dedicated to the prediction of the codeletion of chromosomes 1p and 19q. In particular, Kim et al. (2023) predicted 1p/19q fold change values from H&E-stained WSIs and compared their CNN-based method to the traditional method of detecting 1p/19q status via FISH[54]. Based on a cohort of 288 patients, whose 1p/19q status were confirmed through next generation sequencing, they ascribed a superior predictive power to their proposed method. They further used their model to classify IDH-mutant gliomas into astrocytomas and oligodendrogliomas and validated all results on an external cohort of 385 patients from TCGA.

As various molecular markers show complex interactions and should not be considered as independent variables (e.g., "essentially all 1p/19q co-deleted tumours are also IDH-mutant, although the converse is not true"[2]), Wang et al. (2023) proposed a multiple-instance learning (MIL)- and ViT-based model for predicting IDH mutation, 1p/19q codeletion, and homozygous deletion of CDKN2A/B as well as the presence of necrosis and microvascular proliferation, while simultaneously recognizing interactions between predicted markers[50]. They validated their system in ablation studies and outperformed popular state-of-the-art MIL frameworks and CNN architectures in all tasks, with improvements in accuracy of up to 6.3%.

## Survival prediction

Accurate prediction of survival and a comprehensive understanding of prognostic factors in gliomas are crucial for appropriate disease management[79]. Median survival in adult-type diffuse gliomas ranges from 17 years in oligodendrogliomas to only 8 months in glioblastomas and generally decreases with older age[1]. Besides type, grade, and age, other prognostic factors include tumor size and location, extent of tumor resection, performance status, cognitive impairment, status of molecular markers such as IDH mutation, 1p/19q codeletion, MGMT promoter methylation and treatment protocols, including temozolomide and radiotherapy[2,79]. Table 7 provides an overview of all corresponding studies grouped according to the investigated aspects of prognostic inference.

Almost all reviewed studies investigated the prediction of overall survival from WSIs of H&E-stained tissue sections in an end-to-end manner. One exception inferred predictions by first classifying vascular endothelial cells and quantifying microvascular hypertrophy and hyperplasia[80] and two studies additionally investigated disease-free survival[81,82]. Most studies predicted risk scores or survival times, and only four studies conducted survival prediction by classifying patients into either low- or high-risk groups[83–85] or four predefined risk groups[86].

Given the high variability of disease manifestation and co-morbidities, as well as the stochastic nature of the immediate cause of death, prognosis can be regarded as the most challenging task. In particular, prediction of overall survival from histology has been considered inherently more challenging than tasks such as subtyping and grading[70,87–90], as complex interactions of heterogeneous visual concepts, such as immune cells in general[91], or more specifically lymphocyte infiltrates[92] in the tumor microenvironment, have been recognized as potentially relevant factors.

Zhu et al. (2017) proposed WSISA, the earliest end-to-end system for survival prediction from WSIs, which leveraged a priorly proposed CNN-based survival model[93] to first compute clusters of survival-discriminate image patches and second aggregate cluster-based features for subsequent risk score regression[90].

Potentially due to the aforementioned complexity, only a minority of the reviewed studies proposed uni-modal approaches based on handcrafted features or weakly-supervised CNNs. The majority of the studies utilized either more intricate architectures, or the integration of additional modalities of information.

To this end, Li et al. (2018) and Chen et al. (2021) set out to capture context-aware representations of histological features by employing graph convolutional neural networks, in which vertices corresponded to patch-level feature vectors and edges were defined by either Euclidean distance between feature vectors or adjacency of patches[94,95]. They both compared their models, DeepGraphSurv and Patch-GCN, respectively, on multiple cancer types, where DeepGraphSurv improved on WSISA by 6.5% and Patch-GCN on DeepGraphSurv by 2.6% in terms of C-index.

Three recent studies leveraged the Transformer architecture for modeling histological patterns across patches, either by employing multiple feature detectors to aggregate distinct morphological patterns[88,96] or by fusing information from multiple magnification levels[87]. All studies reported state-of-the-art performance on multiple cancer types, including gliomas, from the TCGA-GBM and TCGA-LGG datasets. Most notably, Liu et al. (2023) reported a 13.2% and a 4.3% C-index performance improvement compared to DeepGraphSurv and Patch-GCN on TCGA-LGG, respectively, and Wang et al. (2023) reported a 5.5%, a 4.5%, and a 3.5% C-index performance improvement compared to DeepGraphSurv, Patch-GCN, and an extension of Patch-GCN via a variance pooling[97] on TCGA-GBM and TCGA-LGG[87,88].

Contrasting the above mentioned uni-modal approaches, half of all studies stated prognosis as a multi-modal problem by integrating clinical or omics data. Simple approaches to such integration included fusing WSI-derived risk scores and variables from other modalities in Cox proportional hazards models[77,82,98] or by concatenating image features and variables prior to subsequent risk score prediction, as proposed by Mobadersany et al. (2018)[40]. In particular, they proposed two CNN-based models, SCNN, which predicted survival from priorly delineated ROIs, and GSCNN, which improved SCNN by concatenating IDH mutation and 1p/19q codeletion status as additional prognostic variables (C-index 0.801 vs. 0.754). They further reported the superiority of concatenation vs. Cox models as strategies for multi-modal fusion (C-index 0.801 vs. 0.757).

More advanced approaches fused distinct uni-modal feature representations into an intermediate feature tensor, using the Kronecker product, while controlling the expressiveness of each modality via a gating-based attention mechanism[70,99,100]. Most notably, by integrating 320 genomics and transcriptomics features in this manner, Chen et al. (2022) surpassed GSCNN on the same ROIs with a C-index performance improvement of 5.8%[70]. The same authors later expanded on this approach by adopting MIL and applying it to 14 cancer types[99]. They conducted extensive analysis regarding the interpretability of their model and reported that the presence of necrosis and IDH1 mutation status were the most attributed features in gliomas from the TCGA-LGG dataset, which is in line with the current WHO classification. They further developed an interactive research tool, to drive the discovery of new prognostic biomarkers.

# Methodological aspects of WSI processing

Across all diagnostic tasks, most of the considered studies processed WSIs in a patch-based manner. The predictive performance of such approaches depends on choices for patch size and magnification, methods for encoding image patches, and approaches for learning relations among patches. In the following, the studies are examined in more detail with regard to these aspects.

## Patch sizes and magnifications

The size of image patches and the magnification at which they are extracted not only determine the number of patches stemming from given WSIs, but also influence the degree to which either cellular details or more distributed characteristics in tissue architecture are captured. However, patch size and magnification were not consistently reported (63/83 and 50/83, respectively) and only three studies compared performance across multiple sizes and magnifications, e.g., patches of size 256 x 256 pixels, and 2.5x, 5x, 10x, and 20x magnification for IDH mutation prediction (AUC 0.80, 0.85, 0.88, and 0.84, respectively)[78]; highest performance in grading by employing 672 x 672 pixels patches at 40x magnification[14]; and highest performance for survival risk group classification consistent across multiple CNN architectures by utilizing 256 x 256 pixels patches at 20x magnification[86]. Apart from these comparisons, smaller patches, i.e., 224 x 224 and 256 x 256 pixels, and 20x magnification were most frequently employed (Figure 4).

In addition to single-magnification approaches, few studies processed multiple magnifications[47,76,78,87,101,102], e.g., by concatenating single-magnification features at the onset[102], averaging of single-magnification predictions[78] or modeling magnifications using graphs[101] or cross-attention[87].

## Encoding of image patches

The reviewed studies employed handcrafted features (13/83), CNNs (63/83) (including ResNet[103], VGG[104], DenseNet[105], EfficientNet[106], Inception[107], and AlexNet[108]), capsule networks (1/83), and ViTs (6/83) to encode image patches. While handcrafted features were mostly employed by earlier approaches, ViT emerged as an alternative to CNNs in recent years. Few studies compared different CNN architectures[36,44,49,75,86] and mostly agreed on the superiority of ResNet over other popular architectures[44,49,75]. Apart from these comparisons, ResNet architectures pre-trained on ImageNet were most commonly employed (33/63) (Figure 5). Self-supervised learning approaches for pre-training CNNs or ViTs using histopathological images leveraged pre-text tasks, including contrastive learning[36,102,109–112], masked pre-training[113] or cross-stain prediction[109].

## Learning paradigms

The majority of all 70 DL-based studies employed either WSL (29/83), MIL (21/83), or previously delineated ROIs (11/83) to process WSIs (the methodological differences of the three approaches are explained in Table 2). In recent years, MIL was most commonly used (Figure 6) and approaches mostly utilized ResNets for encoding image patches (14/21) as well as intricate architectures based on graphs[94,95], Transformers[45,50,87–89,96,99,113,114] or other attention mechanisms[72,83,101,109–111,115] (often combinations thereof) to model histological patterns among patches.

Few studies compared approaches[43,45,50]: Li et al. (2023) conducted insightful comparisons across three subtyping tasks and reported an overall superiority of ViT-L-16 over ResNet50 (both pre-trained on ImageNet) and a worse performance of WSL in all three tasks, with an especially large performance gap in a more fine-grained subtyping task including 11 brain tumor types[45]. Wang et al. (2023), however, reported that their clustering-based CNN model outperformed four state-of-the-art MIL frameworks in several subtyping and grading tasks[43].

# Discussion

As described in the previous sections, a large body of research on the application of AI to various aspects of glioma tissue assessment has been published in recent years. Considering the difficulty of the diagnostic tasks, each study provides important insights for the future application of AI-based methods in real clinical settings. However, there remain serious limitations.

## Limitations of current research

Only 13 of the included studies evaluated the performance of their proposed method on independent test datasets from external pathology departments that were not used during the development of the method[41–43,45,46,54,61,68,72,78,81,86,116]; 10 of these studies were published in 2023 or 2024. External validation, however, is crucial for meaningful performance estimates[117]. Moreover, the performance metrics reported in the solely TCGA-based studies might be inflated, as none of these studies ensured that the data used for model development and performance estimation originated from distinct tissue source-sites[118,119].

As the TCGA datasets were originally compiled for different purposes[37–39], it must be assumed that they are not representative of the intended use of the proposed analysis methods and contain biases. Given these limitations, the performance metrics reported by the reviewed studies should be interpreted with caution and should not be generalized to practical applications[120].

Relatedly, all TCGA-based studies utilized different subpopulations of the two TCGA cohorts without specifying which selection criteria were employed. When studies used other datasets, these were generally not made publicly available. This makes it very hard to reproduce the results and compare results between studies.

Another important aspect is the repeated revision of the WHO classification system, which makes it necessary to revise the now outdated ground truth labels in the TCGA datasets with the current subtypes and grades.

Only three studies utilized the current version from 2021[41–43]; two other studies made an effort to reclassify the TCGA cohorts according to the current 2021 criteria based on available molecular information[49,50], and another three studies considered the previous version from 2016[44–46,101]. All other studies were either based on versions from 2007 or earlier, or did not report the version used, which seriously limits their comparability with the current diagnostic standards, and largely disregards important advances in the understanding of glioma tumor biology. Notably, this concerns important advances regarding the molecular pathogenesis of gliomas with an impact on classification and prognostication, as reflected by the revisions from 2016 and 2021[6,7].

A commonality of many studies in the field is their emphasis on technical aspects of AI approaches, with less focus on the clinical applicability of their proposed methods. The design of several studies suggests that the investigations were primarily based on the availability of data, while a clearly defined clinically relevant research question was not immediately apparent. Examples include studies that mixed distinct glioma types and grades without properly substantiating the motivation for their respective approaches, for instance, when differentiating patients in the TCGA-GBM cohort from those in the TCGA-LGG cohort[15,47,59–62,67,113,121,122]. Moreover, not all studies accurately specified if FFPE or frozen sections were considered for histological assessment. In many studies, it remained unclear for which specific practical diagnostic tasks the proposed methods were envisioned to be relevant.

# Future directions

Considering that the majority of currently available research results are based largely on data from the public TCGA-GBM and TCGA-LGG projects, the acquisition of new datasets with high-quality and up-to-date clinical and molecular pathology annotations is of utmost importance to progress toward clinical applicability. Such new datasets should address the aforementioned limitations, that is, they should consider the latest version of the WHO classification system, be representative of the intended use[120], and include multiple tissue source sites to enable independent external validation[117].

Future work should focus on applications of AI that are clinically relevant, but have received little attention to date. Analysis of intraoperative frozen sections may be valuable for clinicians to improve the robustness of preliminary diagnoses and guide the surgery[42,46,123]. Similarly unexplored is the automated preselection of representative tissue areas for DNA-extraction to speed up the molecular pathology workflow. In general, future work should incorporate a wider variety of staining techniques, including IHC, and further explore their value for AI-based approaches to improve diagnostic tasks or to support predictive and/or prognostic biomarker analysis. In particular, the automated quantification of relevant IHC markers, such as Ki-67 and TP53, and the assessment of their spatial heterogeneity has not yet been investigated specifically for glioma tissue. Besides these applications, future investigations should also consider glioma types other than adult-type diffuse glioma[6].

While the current research applies a wide range of AI methods, some promising approaches remain largely unexplored. The emergence of foundation models for computational pathology[124,125], developed using large amounts of histopathological data, may be an improvement over the current reliance on the ResNet CNN architecture and the ImageNet dataset for pre-training. With their multi-task capabilities, these models could unify many of the reviewed approaches which usually include a specific model for each particular diagnostic task, and potentially open avenues for new clinical use cases[126]. Generative AI models allow for the synthesis of novel histopathological image data, providing new opportunities for AI-based WSI analysis. One of the reviewed studies predicting IDH mutation status has used generative models to synthesize additional training data[75]. Their positive results encourage further investigation. Other promising applications of generative models such as the normalization of staining variations resulting from different image acquisition or tissue staining protocols[127,128], or the synthesis of virtual stainings[129], have not yet been explored in the context of glioma diagnosis.

Most current DL-based methods for WSI analysis predict endpoints directly from pixel data in an end-to-end manner. However, methods considering intermediate, human-understandable features, for example of segmented tissue structures and their spatial relations, can greatly facilitate the verification and communication of results and should remain an active area of work[41,50,55]. Deep learning can also be used to discover such features and reliably segment relevant structures[13,130].

Regarding multi-modal fusion, all studies, across all diagnostic tasks, reported better predictive performance when clinical data, omics data, MRI, and WSIs were used in combination rather than individually. So far, the additional value of WSIs for multi-modal predictive performance has been consistently low[70,82,83,99,131]. Moreover, good predictive performance can often be achieved with much simpler models. For instance, Cox proportional hazards or logistic regression models based on a few clinical variables, such as patient age and sex, can perform on par with WSI-based predictions for survival or IDH mutation status[40,70,77,98,99,132]. Further research should therefore be conducted to better understand and improve the added value of AI-based analysis of WSIs for glioma.

Above all, the reviewed literature clearly indicates a need for future studies with a stronger focus on clinically relevant scientific questions. This requires a truly interdisciplinary approach from the outset, involving neuropathology, neuro-oncology, and the broad expertise of computational science and medical statistics.

# Methods

The literature search for this review was last updated on March 18, 2024 and conducted as follows: The query "(glioma OR glioblastoma OR brain tumor) AND (computational pathology OR machine learning OR deep learning OR artificial intelligence) AND (whole slide image OR WSI)" was used to search PubMed, Web of Science, and Google Scholar. The results from all three searches were then screened for inclusion. Results met the inclusion criteria, if they were original research articles that proposed at least one AI-based method (including traditional ML and DL) for at least one of the diagnostic tasks of subtyping, grading, molecular marker prediction, or survival prediction from whole-slide histopathology images of human gliomas. Studies which incorporated other modalities, e.g., MRI, omics or clinical data, in addition to whole-slide histopathology images were also included. Results were excluded if they were not original research articles (e.g., reviews or book chapters), did not examine glioma or whole-slide histopathology images, primarily investigated image segmentation or retrieval methods, focused on tumor heterogeneity, were conference papers or preprints that were subsequently published as journal articles (in these cases, only journal articles were included), or were not written in English.

The PubMed search yielded exactly 100 results; the search on Web of Science yielded exactly 90 results; and the search on Google Scholar yielded approximately 23,300 results. After screening the first 191 results from Google Scholar (using the setting “Sort by relevance”), no further relevant results were identified. For this reason, it was decided to stop screening after the 300th result and concluded that all further results would likely be irrelevant. From these combined 490 results, 96 duplicates were removed and the remaining 394 results were screened for inclusion. Additional studies that were identified through references or recommendations were also screened for inclusion. As a result, 83 studies, comprising journal articles, conference papers, and preprints, were included in this review.

# Acknowledgements

Research reported in this publication was supported by the German Federal Ministry of Health based on a resolution of the German Bundestag (funding codes: ZMI5-2522DAT15A, ZMI5-2522DAT15B, ZMI5-2522DAT15C, ZMI5-2522DAT15D, ZMI5-2522DAT15E). FF received additional funding from ERACoSysMed and the German Federal Ministry of Education and Research (BMBF) under grant number FKZ 31L0237A (MiEDGE).

# Data Availability Statement

The authors declare that the main data supporting the findings of this study are available within this manuscript.

# Competing Interests

The authors declare no competing interests.

# Author Contributions

JPR and AH conceived the manuscript. JPR conducted the literature review and wrote the manuscript, with feedback from AH, FF, JW, NSS, STH, CB, SL, and AE. The final version of this manuscript was reviewed and approved by all authors.

# Figure and Table Legends

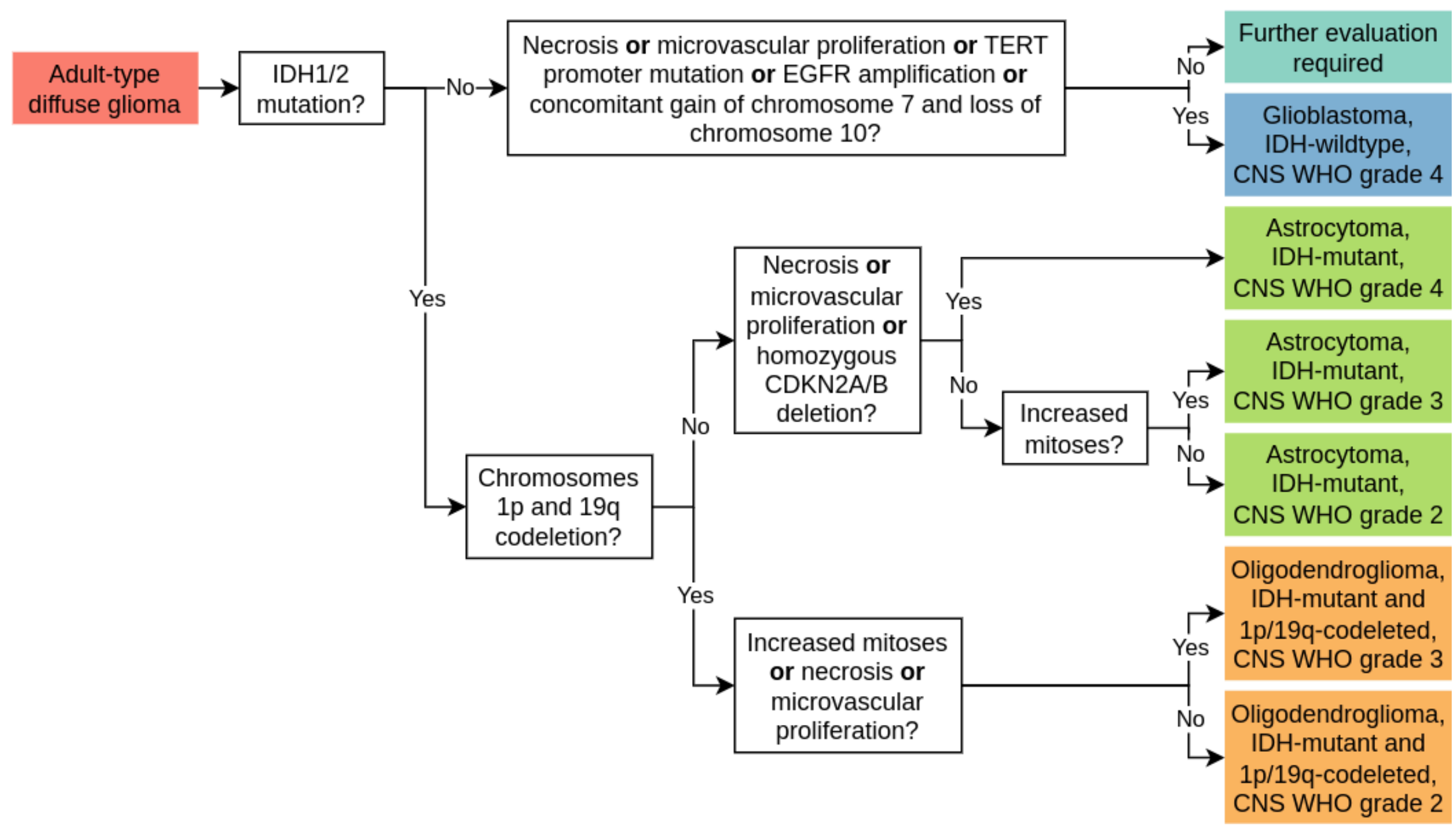

**Figure 1: Schematic diagnostic criteria for adult-type diffuse glioma as defined by the 2021 WHO Classification of Tumors of the Central Nervous System.** The 2021 WHO classification system reorganizes gliomas into adult-type diffuse gliomas, pediatric-type diffuse low-grade and high-grade gliomas, circumscribed astrocytic gliomas, and ependymal tumors[6]. Adult-type diffuse gliomas comprise Glioblastoma, IDH-wildtype, CNS WHO grade 4, Astrocytoma, IDH-mutant, CNS WHO grade 2–4, and Oligodendroglioma, IDH-mutant and 1p/19q-codeleted, CNS WHO grade 2–3[6]. Identified adult-type diffuse glioma are stratified from left to right according to molecular alterations, including mutations in isocitrate dehydrogenase 1/2 (IDH1/2) genes and whole-arm codeletion of chromosomes 1p and 19q, as well as histological features, including increased mitoses, necrosis and/or microvascular proliferation. “or” is understood to be non-exclusive.

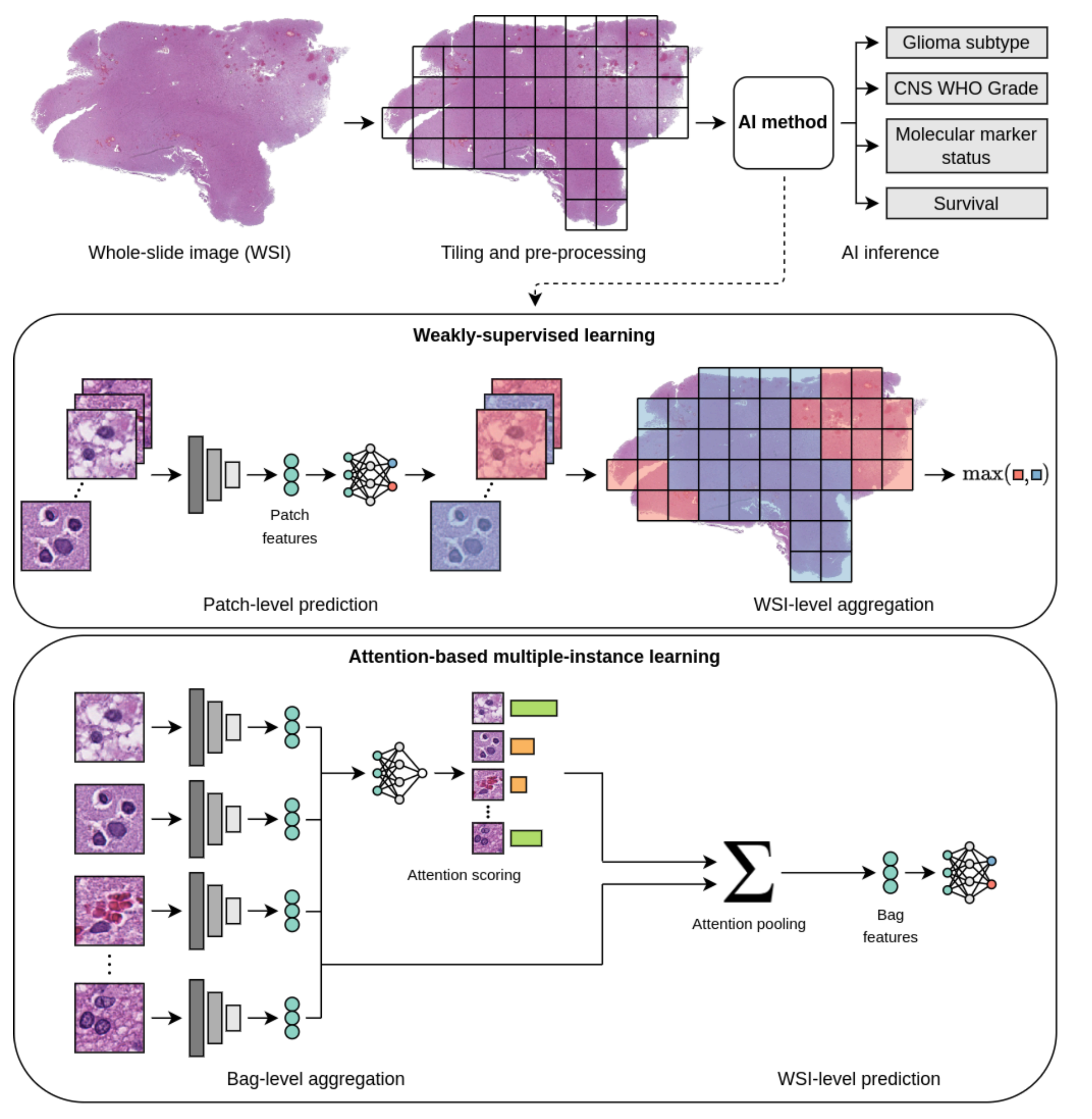

**Figure 2: Typical workflow for artificial intelligence-based analysis of WSIs.** Most approaches partition WSIs into smaller image patches ("Tiling"). Using AI methods, the patches are processed individually and then aggregated to obtain a prediction for the entire WSI. AI methods are often based on weakly-supervised learning or attention-based multiple-instance learning[137], each of which are depicted in greater detail.

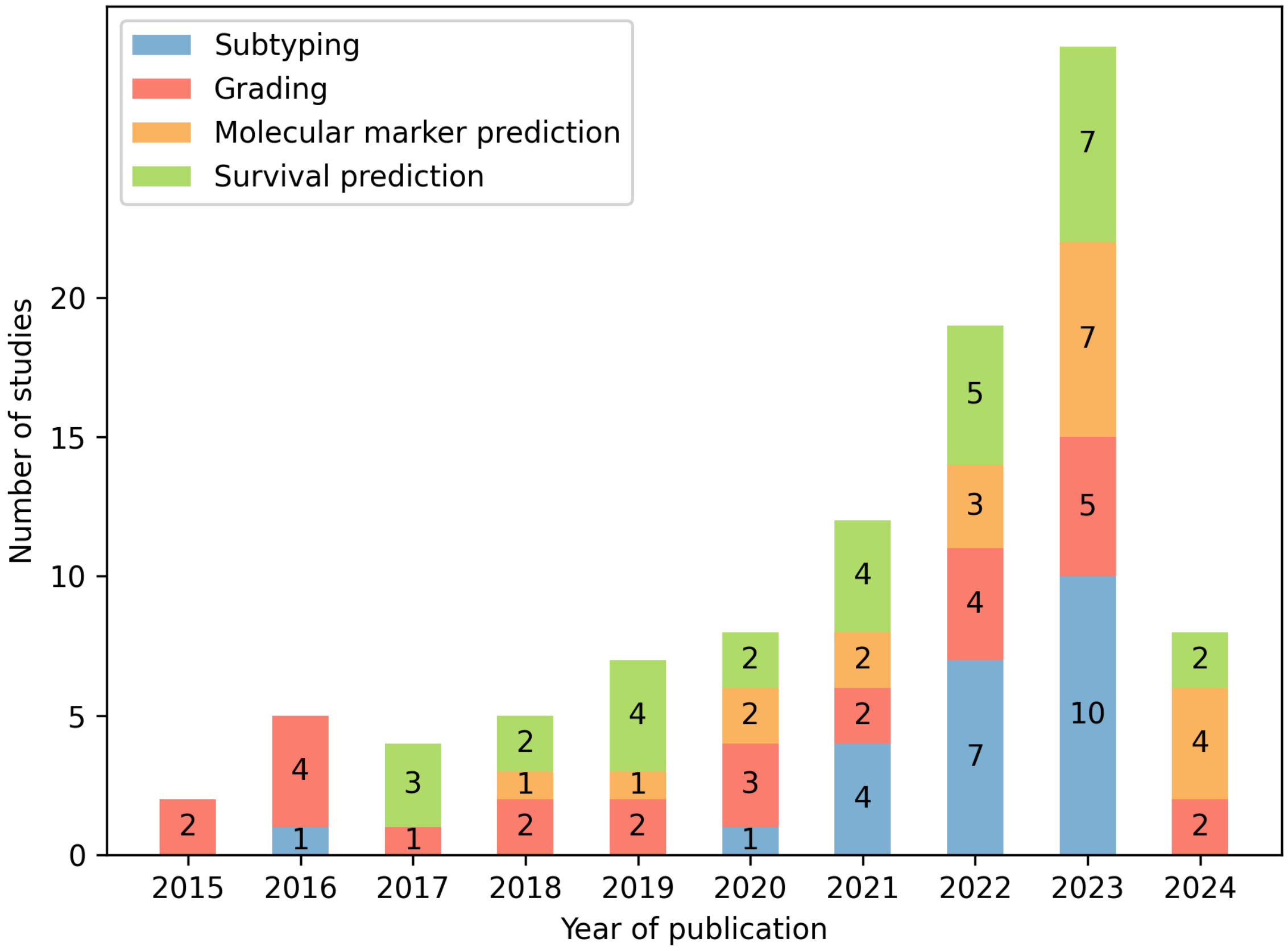

**Figure 3: Number of studies by year of publication and diagnostic task.** All 70 studies included in this review are shown. Studies published in 2024 are considered up until March 18, 2024.

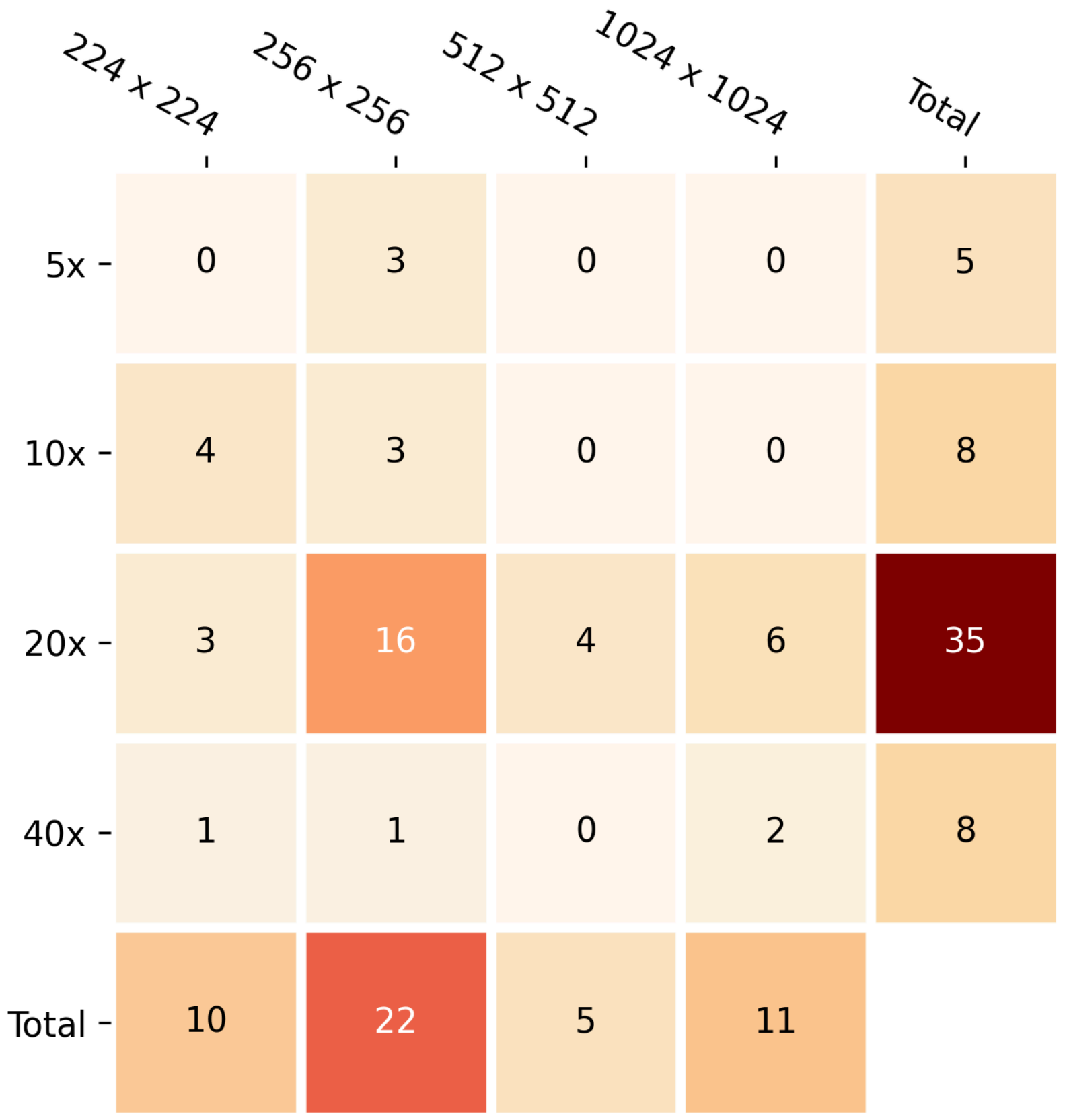

**Figure 4: Patch sizes and magnifications employed by studies.** Studies were taken into account if at least one of the two pieces of information was specified. Other patch sizes and magnifications employed by single studies (e.g., 150 x 150 pixels, 4x magnification) are not shown.

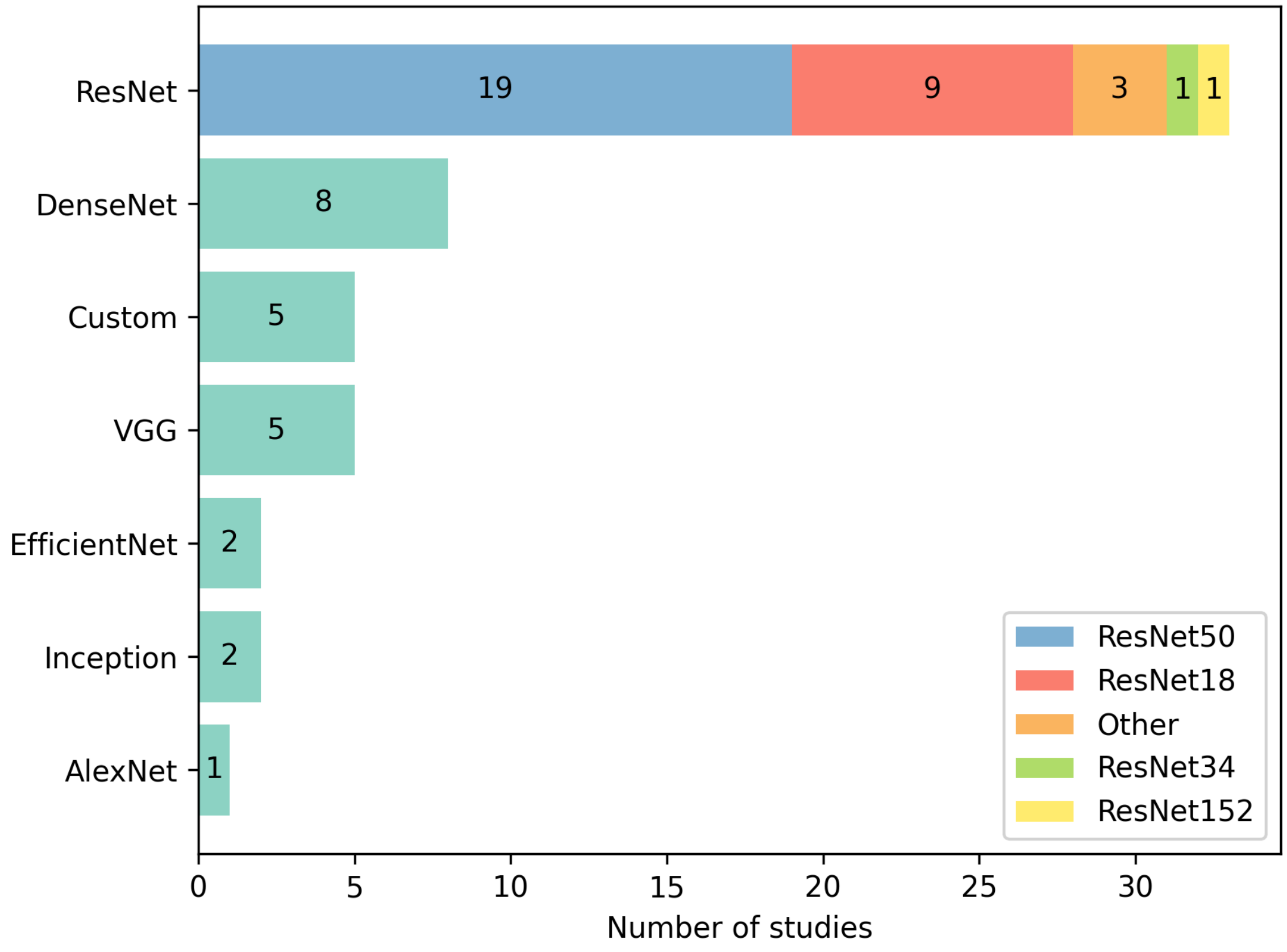

**Figure 5: Convolutional neural network architectures employed by studies.** With only a few exceptions all convolutional neural networks were pre-trained using the ImageNet dataset. Except for ResNet architectures exact variants of stated architectures are not shown. "Custom" refers to custom (i.e., self-configured) architectures.

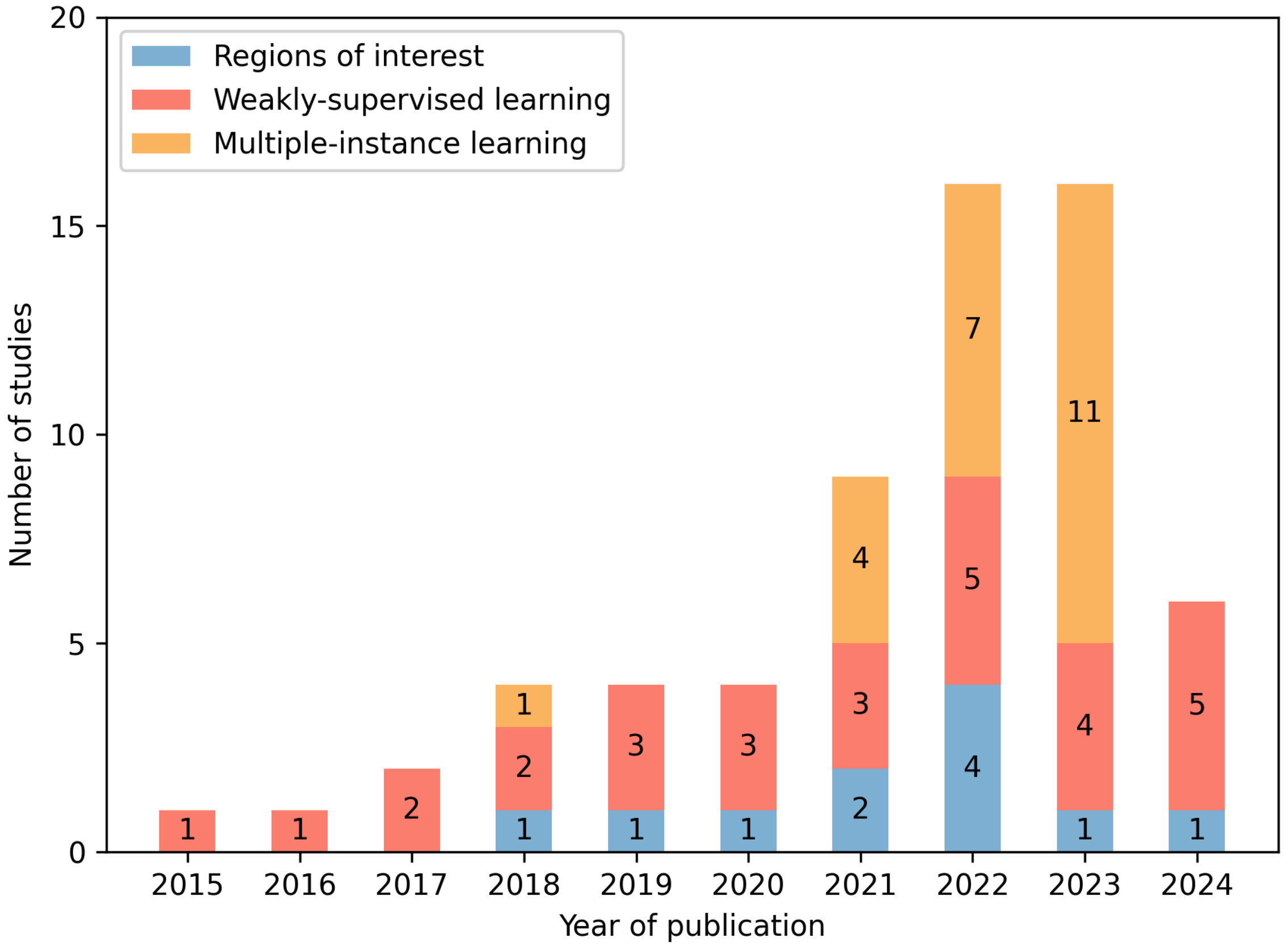

**Figure 6: Learning paradigms employed by studies.** Usage of regions of interest, weakly-supervised learning and multiple-instance learning by all 70 deep learning-based studies included in this review distributed by year of publication. The methodological differences of the three approaches are explained in Table 2.

| | |
|---|---|
| **Artificial intelligence (AI)** | The broad field of AI encompasses the research and development of computer systems that are capable of solving complex tasks typically attributed to human intelligence. |
| **Machine learning (ML)** | ML is a subfield of AI that uses algorithms that automatically learn to recognize patterns from data. In this way, computer systems can be enabled to perform complex cognitive tasks that are difficult to express in the form of human-understandable rules. ML algorithms are mainly used for tasks such as classification (e.g., predicting tumor types), regression (e.g., predicting survival) or segmentation. |
| **Training and Testing** | The development of ML models is typically split into a training and testing phase. During training, an ML algorithm deduces a ML model from a given training dataset. During testing, the model's performance is assessed using a test dataset. The test dataset comprises data which the model has not seen during training, which is crucial for meaningful performance estimates. |
| **Performance metrics** | The performance of ML models is measured by metrics such as accuracy (ACC) or area under the operator receiver characteristic curve (AUC) for classification tasks (e.g., subtyping, grading, molecular marker prediction) or concordance index (C-index)[32] for survival prediction. Metrics closer to 1.0 (resp. 0.5) indicate higher (resp. random) predictive performance. |
| **Features** | To apply ML algorithms to image data, images are usually encoded into a set of features. Traditionally, these features were "handcrafted" by human researchers based on prior knowledge or assumptions to quantify relevant pixel patterns or the morphology or spatial distribution of segmented cell or tissue structures. |
| **Deep learning (DL)** | DL is an advanced form of ML in which deep artificial neural networks are used to automatically learn features from the training data. In this way, more complex cognitive tasks can often be solved than is possible with handcrafted features. Important types of neural network, so called "architectures", for DL-based image analysis are convolutional neural networks (CNNs)[103–108] and Vision Transformers (ViTs)[133]. The training of DL models usually requires very large datasets, which can be mitigated by techniques such as transfer learning[14,134] or self-supervised learning (SSL)[109,135]. |

**Table 1: Essential concepts in artificial intelligence.**

| | |
|---|---|
| **Whole-slide images (WSIs)** | WSIs are digitized microscopic images of whole tissue sections. As they are scanned with a very high resolution of up to 0.25 micrometers per pixel, WSIs can be several gigapixels in size. To enable faster access, WSIs are usually stored as image pyramids consisting of several magnification levels, which are labeled according to the comparable objective magnifications of an analog microscope (e.g., 5x, 10x, 20x, or 40x). The large size and the great variability and complexity of depicted morphological patterns make the analysis of WSI very challenging. |
| **Regions of interest (ROIs)** | As tumor tissue sections naturally include different tissue types, some ML models require a delineation of ROIs of relevant tumor tissue. Usually, this delineation is a laborious task that requires expert knowledge. |
| **Patch-based processing of WSIs** | Due to their gigapixel size, WSIs and ROIs are usually processed by partitioning them into smaller image patches (or tiles) and processing patches individually. This greatly reduces the memory requirements and facilitates parallelization. |
| **Weakly-supervised learning (WSL)** | WSL enables the training of ML models on entire WSIs, eliminating the need for annotating ROIs. The classical approach to WSL assumes that all patches inherit the label (e.g. the tumor type) of the respective WSI. Using these pairs of patches and patch-level labels the ML model is trained to predict a label for each patch. The patch-level predictions are then aggregated (e.g. by averaging) to obtain a WSI-level prediction[136] (Figure 2). |
| **Multiple-instance learning (MIL)** | MIL is a special type of WSL and thus also functions without ROIs. A common approach to MIL groups all patches from a WSI in a “bag” and assigns the label of the WSI to the bag instead of to each individual patch. The ML model is thus trained to predict labels for bags instead of individual patches, enabling the modeling of complex interactions between patches[136]. Examples of MIL include attention-based MIL[137] (Figure 2), and clustering-constrained attention MIL[138]. In the rest of this paper, the term “WSL” only refers to the aforementioned classical WSL approach. |
| **Multi-modal fusion** | Fusing WSIs with other modalities of potentially complementary information, such as MRI, clinical, or omics data, poses great opportunities for improving diagnostic and prognostic precision[139]. Common approaches can be differentiated by the extent to which given modalities are processed in combination before predictions are inferred: Early fusion concatenates modalities at the onset; late fusion aggregates individual uni-modal predictions, e.g., by averaging; and intermediate fusion infers final predictions from intermediate multi-modal representations[139,140]. |

**Table 2: Essential concepts of artificial intelligence-based analysis of WSIs.**

| Dataset | Tumor types | WHO classification version | #Patients | #WSIs | Tissue preparation | Tissue staining | Other information |
|---|---|---|---|---|---|---|---|
| CPTAC-GBM[141] | Glioblastoma | n.s. | 178 | 462 | n.s. | H&E | CT, MRI, omics, clinical |
| Digital Brain Tumor Atlas[142] | 126 distinct brain tumor types | 2016 | 2,880 | 3,115[a] | FFPE | H&E | Clinical |
| Ivy Glioblastoma Atlas Project[143] | Glioblastoma | n.s. | 41 | ~34,500 | Frozen | ISH, H&E | CT, MRI, omics, clinical, annotations |
| TCGA-GBM[37,38] | Glioblastoma | 2000, 2007 | 617 | 2,053 | FFPE, frozen | H&E | Omics, clinical |
| TCGA-LGG[39] | Astrocytoma, anaplastic astrocytoma, oligodendroglioma, oligoastrocytoma, anaplastic oligoastrocytoma | 2000, 2007 | 516 | 1,572 | FFPE, frozen | H&E | Omics, clinical |
| UPENN-GBM[144] | Glioblastoma | 2016 | 34 | 71 | n.s. | H&E | MRI, omics, clinical |

**Table 3: Overview of all publicly available datasets utilized by reviewed studies.** Other datasets that were utilized but were not made publicly available are not listed. FFPE = formalin-fixed and paraffin-embedded, H&E = hematoxylin and eosin staining, ISH = in situ hybridization, CT = computed tomography, MRI = magnetic resonance imaging, n.s. = not specified, [a] = including 47 non-tumor WSIs

| Study | Method | | Multi-modal fusion | | Dataset | | | | | Performance | |
|---|---|---|---|---|---|---|---|---|---|---|---|
| | Features | Learning paradigm | Modalities | Fusion Strategy | TCGA-GBM | TCGA-LGG | Other | #Patients | #WSIs | Metric | Result |
| **Astrocytoma vs. Oligodendroglioma** | | | | | | | | | | | |
| Jin et al. (2023)[56] | CNN (DenseNet) | SL | H&E | | | | X | n.s. | 733 | ACC | 0.920 |
| Kurc et al. (2020)[13] | CNN (DenseNet) | WSL | H&E, MRI | Late | | X | | 52 | n.s. | ACC | 0.900 |
| **Astrocytoma IDH-mutant vs. Oligodendroglioma IDH-mutant, 1p/19q codeleted** | | | | | | | | | | | |
| Kim et al. (2023)[54] | CNN (RetCCL) | WSL | H&E | | X | X | X | 673 | 673 | AUC | 0.837 |
| **Astrocytoma IDH-mutant vs. Astrocytoma IDH-wildtype vs. Oligodendroglioma IDH-mutant, 1p/19q codeleted** | | | | | | | | | | | |
| Faust et al. (2022)[35] | CNN (VGG19) | SL | H&E, IHC | Late | | | X | n.s. | 1,013 | ACC | 1.0 |
| **Astrocytoma IDH-mutant vs. Oligodendroglioma IDH-mutant, 1p/19q codeleted vs. Glioblastoma IDH-wildtype** | | | | | | | | | | | |
| Hewitt et al. (2023)[41] | CNN, ViT (CTransPath) | MIL | H&E | | X | X | X | 2,741 | n.s. | AUC | 0.840[a], 0.910[b], 0.900[c] |
| Nasrallah et al. (2023)[42] | ViT | WSL | IFS | | X | X | X | 1,524 | 2,334 | AUC | 0.900[a], 0.880[b], 0.930[c] |
| Wang et al. (2023)[43] | CNN (ResNet50) | WSL | H&E | | | | X | 2,624 | 2,624 | AUC | 0.941[a], 0.973[b], 0.983[c] |
| Wang et al. (2023)[50] | ViT | MIL | H&E | | X | X | | 940 | 2,633 | ACC | 0.773 |
| Jose et al. (2022)[49] | CNN (ResNet50) | WSL | H&E | | X | X | | 700 | 926 | AUC | 0.961 |
| **Astrocytoma vs. Oligodendroglioma vs. Glioblastoma** | | | | | | | | | | | |
| Hsu et al. (2022)[51] | CNN (ResNet50) | WSL | H&E, MRI | Late | X | X | X | 329 | n.s. | balanced ACC | 0.654 |
| Mallya et al. (2022)[53] | CNN | MIL | H&E, MRI | KD | | | X | n.s. | 221 | balanced ACC | 0.752 |
| Suman et al. (2022)[115] | CNN | WSL, MIL | H&E | | X | X | | 230 | n.s. | balanced ACC | 0.730 |
| Wang et al. (2022)[52] | CNN (EfficientNet, SEResNeXt101) | WSL | H&E, MRI, clinical | Late | X | X | X | 378 | n.s. | balanced ACC | 0.889 |
| Lu et al. (2021)[110] | CNN (ResNet50) | SSL, MIL | H&E | Intermediate | X | X | | 700 | 700 | ACC | 0.886 |
| **Astrocytoma IDH-mutant vs. Astrocytoma IDH-wildtype vs. Oligodendroglioma IDH-wildtype vs. Glioblastoma IDH-mutant vs. Glioblastoma IDH-wildtype** | | | | | | | | | | | |
| Chitnis et al. (2023)[114] | CNN (KimiaNet) | MIL | H&E | | | | X | 791 | 866 | AUC | 0.969 |
| **Astrocytoma vs. Anaplastic Astrocytoma vs. Oligodendroglioma vs. Anaplastic Oligodendroglioma vs. Glioblastoma** | | | | | | | | | | | |
| Li et al. (2023)[45] | ViT (ViT-L-16) | MIL | H&E | | X | X | X | 749 | 1,487 | AUC | 0.932 |
| Jin et al. (2021)[48] | CNN (DenseNet) | WSL | H&E | | | | X | 323 | 323 | ACC | 0.865 |
| Xing et al. (2021)[101] | CNN (DenseNet) | WSL | H&E | | | | X | n.s. | 440 | ACC | 0.700 |
| **Astrocytoma vs. Anaplastic Astrocytoma vs. Oligodendroglioma vs. Anaplastic Oligodendroglioma vs. Oligoastrocytoma vs. Glioblastoma** | | | | | | | | | | | |
| Hou et al. (2016)[47] | CNN | WSL | H&E | | X | X | | 539 | 1,064 | ACC | 0.771 |
| **Astrocytoma vs. Oligodendroglioma vs. Intracranial Germinoma** | | | | | | | | | | | |
| Shi et al. (2023)[46] | CNN (ResNet152) | WSL | H&E, IFS | | | X | X | 346 | 832 | ACC | 0.769, 0.820[d] |
| **Astrocytoma vs. Oligodendroglioma vs. Ependymoma vs. Lymphoma vs. Metastasis vs. Non-tumor** | | | | | | | | | | | |
| Ma et al. (2023)[55] | CNN (DenseNet) | SL | H&E | | | | X | n.s. | 1,038 | ACC | 0.935 |
| **Astrocytic vs. Oligodendroglial vs. Ependymal** | | | | | | | | | | | |
| Yang et al. (2022)[109] | CNN (ResNet18) | SSL, MIL | H&E | | | | X | n.s. | 935 | weighted F1-score | 0.780 |
| **Oligodendroglioma vs. Non-Oligodendroglioma** | | | | | | | | | | | |
| Im et al. (2021)[44] | CNN (ResNet50) | ROIs | H&E | | | | X | 369 | n.s. | balanced ACC | 0.873 |

**Table 4: Overview of all studies related to subtyping.** Studies are organized into specific subtasks and sorted by year of publication. Columns “TCGA-GBM”, “TCGA-LGG” and “Other” indicate whether TCGA-GBM, TCGA-LGG and/or other public or proprietary datasets were employed. “#Patients” and “#WSIs” state the reported size of the dataset including subsets for training and validation. “Performance”

states a subset of the reported performance metrics. CNN = convolutional neural network, ViT = Vision Transformer, ROI = region of interest, SL = supervised learning, WSL = weakly-supervised learning, SSL = self-supervised learning, MIL = multiple-instance learning, KD = knowledge distillation, H&E = hematoxylin and eosin staining, MRI = magnetic resonance imaging, IHC = immunohistochemistry, IFS = intraoperative frozen sections, ACC = accuracy, AUC = area under the operator receiver curve, n.s. = not specified, [a] = for astrocytoma, [b] = for oligodendroglioma, [c] = for glioblastoma, [d] = for model based on intraoperative frozen sections.

| Study | Method | | Multi-modal fusion | | Dataset | | | | | Performance | |
|---|---|---|---|---|---|---|---|---|---|---|---|
| | Features | Learning paradigm | Modalities | Fusion Strategy | TCGA-GBM | TCGA-LGG | Other | #Patients | #WSIs | Metric | Result |
| **WHO grade II vs. WHO grade III** | | | | | | | | | | | |
| Su et al. (2023)[66] | CNN (ResNet18) | WSL | H&E | | | X | | 507 | n.s. | ACC | 0.801 |
| **CNS WHO grade 2 vs. CNS WHO grade 3 vs. CNS WHO grade 4** | | | | | | | | | | | |
| Jin et al. (2023)[56] | CNN (DenseNet) | SL | H&E | | | | X | n.s. | 733 | ACC | 0.730 |
| Wang et al. (2023)[43] | CNN (ResNet50) | WSL | H&E | | | | X | 2,624 | 2,624 | AUC | 0.939[a], 0.930[b], 0.990[c], 0.948[d], 0.967[e] |
| **WHO grade II vs. WHO grade III vs. WHO grade IV** | | | | | | | | | | | |
| Qiu et al. (2023)[69] | CNN (ResNet) | ROIs | H&E, omics | Intermediate | X | X | | 683 | 683 | AUC | 0.872 |
| Zhang et al. (2022)[111] | CNN (EfficientNet) | SSL, MIL | H&E | | X | X | | 499 | n.s. | ACC | 0.790[f], 0.750[g] |
| Xing et al. (2022)[71] | CNN (ResNet18) | ROIs | H&E, omics | KD | X | X | | 736 | 1,325* | AUC | 0.924 |
| Pei et al. (2021)[63] | CNN (ResNet) | WSL | H&E, omics | Early | X | X | | n.s. | 549 | ACC | 0.938[h], 0.740[i] |
| Chen et al. (2022)[70] | CNN (VGG19) | ROIs | H&E, omics | Intermediate | X | X | | 769 | 1,505* | AUC | 0.908 |
| Truong et al. (2020)[64] | CNN (ResNet18) | WSL | H&E | | X | X | | 1,120 | 3,611 | ACC | 0.730[h], 0.530[i] |
| Wang et al. (2019)[58] | Handcrafted | | H&E, PI | Early | | | X | 146 | n.s. | ACC | 0.900 |
| Ertosun et al. (2015)[65] | CNN | WSL | H&E | | X | X | | n.s. | 37 | ACC | 0.960[h], 0.710[i] |
| **WHO grade II and III vs. WHO grade IV** | | | | | | | | | | | |
| Jiang et al. (2023)[113] | CNN (ResNet18) | SSL, MIL | H&E | | X | X | | n.s. | n.s. | AUC | 0.964 |
| Brindha et al. (2022)[121] | Handcrafted | | H&E | | X | X | | n.s. | 1,114 | ACC | 0.972 |
| Mohan et al. (2022)[59] | Handcrafted | | H&E | | X | X | | n.s. | 310 | AUC | 0.974 |
| Im et al. (2021)[44] | CNN (MnasNet) | ROIs | H&E | | | | X | 468 | n.s. | balanced ACC | 0.580 |
| Rathore et al. (2020)[60] | Handcrafted | ROIs | H&E, clinical | Early | X | X | | 735 | n.s. | AUC | 0.927 |
| Momeni et al. (2018)[67] | CNN | WSL | H&E | | X | X | | 710 | n.s. | AUC | 0.930 |
| Yonekura et al. (2018)[122] | CNN | WSL | H&E | | X | X | | n.s. | 200 | ACC | 0.965 |
| Xu et al. (2017)[14] | CNN (AlexNet) | WSL | H&E | | | | X | n.s. | 85 | ACC | 0.975 |
| Barker et al. (2016)[61] | Handcrafted | | H&E | | X | X | X | n.s. | 649 | AUC | 0.960 |
| Hou et al. (2016)[47] | CNN | WSL | H&E | | X | X | | 539 | 1,064 | ACC | 0.970 |
| Reza et al. (2016)[62] | Handcrafted | | H&E | | X | X | | n.s. | 66 | AUC | 0.955 |
| Fukuma et al. (2016)[15] | Handcrafted | | H&E | | X | X | | n.s. | 300 | ACC | 0.996 |
| **WHO grade I vs. WHO grade II vs. WHO grade III vs. WHO grade IV** | | | | | | | | | | | |
| Elazab et al. (2024)[145] | CNN (ResNet50) | WSL | H&E | | X | X | | 445 | 654 | ACC | 0.972 |
| **WHO grade I and II vs. WHO grade III and IV** | | | | | | | | | | | |
| Mousavi et al. (2015)[57] | Handcrafted | | H&E | | X | X | | 138 | n.s. | ACC | 0.847 |
| **Non-tumor vs. WHO grade I vs. WHO grade II vs. WHO grade III vs. WHO grade IV** | | | | | | | | | | | |
| Pytlarz et al. (2024) | CNN (DenseNet) | SSL, WSL | HLA | | | | X | n.s. | 206*** | ACC | 0.690 |
| **Non-tumor vs. WHO grade II and III vs. WHO grade IV** | | | | | | | | | | | |
| Ker et al. (2019)[134] | CNN (InceptionV3) | WSL | H&E | | | | X | n.s. | 154** | F1-score | 0.991 |

**Table 5: Overview of all studies related to grading.** Studies are organized into specific subtasks and sorted by year of publication. Columns “TCGA-GBM”, “TCGA-LGG” and “Other” indicate whether TCGA-GBM, TCGA-LGG and/or other public or proprietary datasets were employed. “#Patients” and “#WSIs” state the reported size of the dataset including subsets for training and validation. “Performance” states a subset of the reported performance metrics. CNN = convolutional neural network, ROI = region of interest, SL = supervised learning, WSL = weakly-supervised learning, SSL = self-supervised learning, MIL = multiple-instance learning, KD = knowledge distillation, H&E =

hematoxylin and eosin staining, HLA = human leukocyte antigen staining, PI = proliferation index, ACC = accuracy, AUC = area under the operator receiver curve, n.s. = not specified, [a/b/c] = for astrocytoma CNS WHO grade 2/3/4, [d/e] = for oligodendroglioma CNS WHO grade 2/3, [f] = for frozen sections, [g] = for FFPE sections, [h] = for II+III vs. IV, [i] = for II vs. III, * = not WSIs but ROIs, ** = probably not gigapixel WSIs but smaller histopathology images, *** = tissue microarrays.

| Study | Method | | Multi-modal fusion | | Dataset | | | | | Performance | |
|---|---|---|---|---|---|---|---|---|---|---|---|
| | Features | Learning paradigm | Modalities | Fusion Strategy | TCGA-GBM | TCGA-LGG | Other | #Patients | #WSIs | Metric | Result |
| **IDH mutation** | | | | | | | | | | | |
| Zhao et al. (2024)[68] | CNN (ResNet50), ViT | WSL | H&E | | | | X | 2,275 | n.s. | AUC | 0.953 |
| Liechty et al. (2022)[78] | CNN (DenseNet) | WSL | H&E | | X | X | X | 513 | 975 | AUC | 0.881 |
| Jiang et al. (2021)[77] | CNN (ResNet18) | WSL | H&E, clinical | Early | | X | | 490 | 843 | AUC | 0.814 |
| Wang et al. (2021)[76] | Handcrafted | | H&E, MRI | Late | X | X | X | 217 | n.s. | ACC | 0.900 |
| Liu et al. (2020)[75] | CNN (ResNet50) | WSL | H&E, clinical | Early | X | X | X | 266 | n.s. | AUC | 0.931 |
| **MGMT promoter methylation** | | | | | | | | | | | |
| He et al. (2024)[116] | ViT (Swin Transformer V2) | ROIs, WSL | H&E | | X | | X | 224 | 657 | AUC | 0.860 |
| Yin et al. (2024)[146] | Spiking neural P system | | H&E | | X | | | 110 | 110 | ACC | 0.700 |
| Dai et al. (2023)[147] | Spiking neural P system | | H&E | | X | | | 110 | 110 | ACC | 0.681 |
| **IDH mutation, MGMT promoter methylation** | | | | | | | | | | | |
| Krebs et al. (2023)[102] | CNN (ResNet18) | SSL, MIL | H&E | | X | | | 325 | 196[f], 216[g] | ACC | 0.912[a], 0.861[b] |
| **IDH mutation, MGMT promoter methylation, TP53 mutation** | | | | | | | | | | | |
| Li et al. (2023)[45] | ViT (ViT-L-16) | MIL | H&E | | X | X | X | 1,005[a], 257[b], 877[c] | n.s. | AUC | 0.960[a], 0.845[b], 0.874[c] |
| **IDH mutation, MGMT promoter methylation, 1p/19q codeletion** | | | | | | | | | | | |
| Momeni et al. (2018)[67] | CNN | WSL | H&E | | X | X | | 710 | n.s. | AUC | 0.860[a], 0.750[b], 0.760[d] |
| **1p/19q fold change values** | | | | | | | | | | | |
| Kim et al. (2023)[54] | CNN (RetCCL) | WSL | H&E | | X | X | X | 673 | 673 | AUC | 0.833[e], 0.837[f] |
| **IDH mutation, 1p/19q codeletion** | | | | | | | | | | | |
| Rathore et al. (2019)[148] | CNN (ResNet) | ROIs | H&E | | X | X | | 663 | n.s. | AUC | 0.860[a], 0.860[d] |
| **IDH mutation, 1p/19q codeletion, homozygous deletion of CDKN2A/B** | | | | | | | | | | | |
| Wang et al. (2023)[50] | ViT | MIL | H&E | | X | X | | 940 | 2,633 | AUC | 0.920[a], 0.881[d], 0.772[g] |
| **IDH mutation, 1p/19q codeletion, homozygous deletion of CDKN2A/B, ATRX mutation, EGFR amplification, TERT promoter mutation** | | | | | | | | | | | |
| Hewitt et al. (2023)[41] | CNN, ViT (CTransPath) | MIL | H&E | | X | X | X | 2,840 | n.s. | AUC | 0.900[a], 0.870[d], 0.730[g], 0.790[h], 0.850[i], 0.600[j] |
| **IDH mutation, 1p/19q codeletion, homozygous deletion of CDKN2A/B, ATRX mutation, EGFR amplification, TP53 mutation, CIC mutation** | | | | | | | | | | | |
| Nasrallah et al. (2023)[42] | ViT | WSL | IFS | | X | X | X | 1,524 | 2,334 | AUC | 0.820[a], 0.820[d], 0.800[g], 0.790[h], 0.710[i], 0.870[c], 0.790[k] |
| **DNA methylation** | | | | | | | | | | | |
| Zheng et al. (2020)[149] | Handcrafted | | H&E, clinical | Early | X | X | | 327 | n.s. | AUC | 0.740 |
| **Pan-cancer studies** | | | | | | | | | | | |
| Saldanha et al. (2023)[72] | CNN (RetCCL) | MIL | H&E | | X | | X | 493 | n.s. | AUC | 0.840[a], 0.700[c], 0.700[h] |
| Arslan et al. (2024)[81] | CNN (ResNet34) | WSL | H&E | | X | X | X | 894 | 1,999 | AUC | 0.669[l] |
| Loeffler et al. (2022)[74] | CNN (ShuffleNet) | ROIs | H&E | | X | X | | 680 | n.s. | AUC | 0.764[a], 0.787[c], 0.726[h] |

**Table 6: Overview of all studies related to molecular marker prediction.** Studies are organized into specific subtasks and sorted by year of publication. Columns "TCGA-GBM", "TCGA-LGG" and "Other" indicate whether TCGA-GBM, TCGA-LGG and/or other public or proprietary

datasets were employed. “#Patients” and “#WSIs” state the reported size of the dataset including subsets for training and validation. “Performance” states a subset of the reported performance metrics. CNN = convolutional neural network, ViT = Vision Transformer,SL = supervised learning, ROI = region of interest, WSL = weakly-supervised learning, SSL = self-supervised learning, MIL = multiple-instance learning, H&E = hematoxylin and eosin staining, IFS = intraoperative frozen sections, MRI = magnetic resonance imaging, ACC = accuracy, AUC = area under the operator receiver curve, n.s. = not specified, [a] = for IDH mutation, [b] = for MGMT promoter methylation, [c] = for TP53 mutation, [d] = for 1p/19q codeletion, [e] = for 1p, [f] = for 19q, [g] = for CDKN2A/B, [h] = for ATRX, [i] = for EGFR, [j] = for TERT, [k] = for CIC, [l] = average performance.

| Study | Method | | Multi-modal fusion | | Dataset | | | | | Performance | |
|---|---|---|---|---|---|---|---|---|---|---|---|
| | Features | Learning paradigm | Modalities | Fusion Strategy | TCGA-GBM | TCGA-LGG | Other | #Patients | #WSIs | Metric | Result |
| **Risk score/survival time regression** | | | | | | | | | | | |
| Yin et al. (2024)[146] | Spiking neural P system | | H&E | | X | | | 110 | 110 | PCC | 0.541 |
| Dai et al. (2023)[147] | Spiking neural P system | | H&E | | X | | | 110 | 110 | PCC | 0.515 |
| Jiang et al. (2023)[96] | CNN (ResNet18) | MIL | H&E | | | X | | 490 | 843 | C-index | 0.714 |
| Jiang et al. (2023)[113] | CNN (ResNet18) | SSL, MIL | H&E | | | X | | n.s. | n.s. | C-index | 0.685 |
| Liu et al. (2023)[87] | CNN (ResNet50) | MIL | H&E | | | X | | 486 | 836 | C-index | 0.702 |
| Luo et al. (2023)[82] | CNN (ResNet50) | SL | H&E, clinical | Early | | | X | 162 | n.s. | C-index | 0.928** |
| Wang et al. (2023)[88] | CNN (ResNet50) | MIL | H&E | | X | X | | 1,041 | n.s. | C-index | 0.861 |
| Carmichael et al. (2022)[97] | CNN (ResNet18) | MIL | H&E | | X | X | | 872 | n.s. | C-index | 0.738 |
| Chen et al. (2022)[70] | CNN (VGG19) | ROIs | H&E, omics | Intermediate | X | X | | 769 | 1,505* | C-index | 0.826 |
| Chunduru et al. (2022)[98] | CNN (ResNet50) | ROIs | H&E, omics, clinical | Late | X | X | | 766 | 1,061 | C-index | 0.840 |
| Braman et al. (2021)[100] | CNN (VGG19) | ROIs | H&E, omics, clinical, MRI | Intermediate | X | X | | 176 | 372* | C-index | 0.788 |
| Chen et al. (2021)[89] | CNN (ResNet50) | MIL | H&E, omics | Intermediate | X | X | | 1,011 | n.s. | C-index | 0.817 |
| Chen et al. (2021)[95] | CNN (ResNet50) | MIL | H&E | | X | X | | 1,011 | n.s. | C-index | 0.824 |
| Jiang et al. (2021)[77] | CNN (ResNet18) | WSL | H&E, omics, clinical | Early | | X | | 490 | 843 | C-index | 0.792 |
| Hao et al. (2019)[131] | CNN | WSL | H&E, omics, clinical | Late | X | | | 447 | n.s. | C-index | 0.702 |
| Rathore et al. (2019)[148] | CNN (ResNet) | ROIs | H&E | | X | X | | 663 | n.s. | C-index | 0.820 |
| Tang et al. (2019)[150] | Capsule network | WSL | H&E | | X | | | 209 | 424 | C-index | 0.670 |
| Li et al. (2018)[94] | CNN (VGG16) | MIL | H&E | | X | | | 365 | 491 | C-index | 0.622 |
| Mobadersany et al. (2018)[40] | CNN (VGG19) | ROIs | H&E, omics | Late | X | X | | 769 | 1,061* | C-index | 0.801 |
| Nalisnik et al. (2017)[80] | Handcrafted | | H&E, clinical | Late | | X | | n.s. | 781 | C-index | 0.780 |
| Zhu et al. (2017)[90] | CNN | WSL | H&E | | X | | | 126 | 255 | C-index | 0.645 |
| **Survival time distribution modeling** | | | | | | | | | | | |
| Liu et al. (2022)[151] | CNN (ResNet50) | MIL | H&E | | | X | | 486 | 836 | C-index | 0.642 |
| **Risk group classification** | | | | | | | | | | | |
| Baheti et al. (2023)[83] | CNN (ResNet50) | MIL | H&E, omics, clinical | Late | X | X | | 188 | 188 | AUC | 0.746 |
| Shirazi et al. (2020)[86] | CNN (Inception) | ROIs | H&E | | X | X | X | 454 | 858* | AUC | 1.0 |
| Zhang et al. (2020)[84] | Handcrafted | | H&E, omics, clinical | Late | X | | | 606 | 1,194 | AUC | 0.932 |
| Powell et al. (2017)[85] | Handcrafted | | H&E, omics, clinical | Early | | X | | 53 | n.s. | AUC | 0.890 |
| **Pan-cancer studies** | | | | | | | | | | | |
| Arslan et al. (2024)[81] | CNN (ResNet34) | WSL | H&E | | X | X | | 705 | 1,489 | n.s. | n.s. |
| Chen et al. (2022)[99] | CNN (ResNet50) | MIL | H&E, omics | Intermediate | | X | | 479 | n.s. | C-index | 0.808 |
| Cheerla et al. (2019)[112] | CNN (SqueezeNet) | SSL | H&E, omics, clinical | Intermediate | | X | | n.s. | n.s. | C-index | 0.850 |

**Table 7: Overview of all studies related to survival prediction.** Studies are organized into specific subtasks and sorted by year of publication. Columns “TCGA-GBM”, “TCGA-LGG” and “Other” indicate whether TCGA-GBM, TCGA-LGG and/or other public or proprietary datasets were employed. “#Patients” and “#WSIs” state the reported size of the dataset including subsets for training and validation. “Performance” states a subset of the reported performance metrics. CNN = convolutional neural network, ROI = region of interest, SL = supervised learning, WSL = weakly-supervised learning, SSL = self-supervised learning, MIL = multiple-instance learning, H&E = hematoxylin and eosin staining, MRI = magnetic resonance imaging, PCC = Pearson correlation coefficient, C-index = concordance index, AUC = area under the operator receiver curve, n.s. = not specified, * = not WSIs but ROIs, ** = probably not gigapixel WSIs but smaller histopathology images.